\documentstyle{laa}
\input psfig
\begin{document}
\newcommand{\proj}[1] {{#1'}}
\newcommand{\projd}[2] {{#1'\!\!_{\mbox{#2}}}}
\newcommand{\conv}[1] {{\overline{#1}}}
\newcommand{\obs}[1] {{#1_{\mbox{obs}}}}
\newcommand{\ave}[1] {\langle{#1}\rangle}
%
\newcommand{\ten}[1] {$10 ^{#1}$}
%
\newcommand{\ML}[1] {$\mbox{M/L}_{\mbox{#1}}$}
%
\newcommand{\kms} {$\mbox{km.s}^{-1}$}
\newcommand{\Msun} {$\mbox{M}_{\sun}$}
\newcommand{\Lsun} {$\mbox{L}_{\sun}$}
\newcommand{\Jansky} {$\mbox{J}_{\mbox{y}}$ }
\newcommand{\micron} {$\mu m$ }
\newcommand{\Kelvin} {\degr K }
\newcommand{\magsec} {$\mbox{mag.arcsec}^{-2}$} 
\newcommand{\dpc} {$\mbox{pc}^{-2}$} 
\newcommand{\Ag} {\AA~} 
\newcommand{\Wm} {$\mbox{W.m}^{-2}$} 
%
\newcommand{\diff} {\mbox{d}}
\newcommand{\dint} {\int\!\!\int}
\newcommand{\tint} {\int\!\!\int\!\!\int}
%
\newcommand{\Halpha} {$\mbox{H}_\alpha$}


\def\spose#1{\hbox to 0pt{#1\hss}}
\def\lta{\mathrel{\spose{\lower 3pt\hbox{$\sim$}}
    \raise 2.0pt\hbox{$<$}}}
\def\gta{\mathrel{\spose{\lower 3pt\hbox{$\sim$}}
    \raise 2.0pt\hbox{$>$}}}

\thesaurus{11(11.09.1 M31; 11.14.1; 11.11.1)}

\title{N body simulations of the nucleus of M31}

\author{E. Emsellem \inst{1,2}
\and F. Combes \inst{3}}

\offprints{E. Emsellem (email: eemselle@eso.org)}

\institute{European Southern Observatory, Karl-Schwarschild Strasse 2, D-85748
Garching b. M\"unchen, Germany \and 
Sterrewacht Leiden, Postbus 9513, 2300 RA Leiden, The Netherlands 
\and Observatoire de Paris, DEMIRM, 61 av. de l'Observatoire, F-75014 Paris}

\date{accepted 14/01, 1997}

\maketitle
\markboth{E. Emsellem \& F. Combes: N body simulations of the nucleus of M 
31}{}

\begin{abstract}
We test through stellar N-body simulations some scenarios to explain
the dynamics of the peculiar nucleus of the Andromeda galaxy (M~31):
although HST observations reveal a double nucleus morphology, 
the rotation field is almost symmetric around the bulge
gravity centre and the velocity dispersion is off-centred. We show that
any $m=1$ perturbation has a very short life-time (a few 10$^5$ yr). 
Assuming that the bright peak (P1) is a cold stellar cluster infalling into 
the nucleus, and that the large central velocity gradient is due to a 
central dark mass (in the range 7~$10^7$--$10^8$~\Msun), we obtain a 
reasonably good fit to the observations.
However, if this cluster lies in the central 20 pc,
we estimate the life-time of the cluster to be less than 0.5~Myr. 
The dynamical friction is more efficient than estimated by analytic 
formulae, and is essentially due to the deformation of the stellar 
cluster through the huge tidal forces provided by the black hole.
We show that the cluster cannot be on a circular orbit
around the centre if the nucleus hosts a massive black hole of
a few $10^7$~\Msun, and finally provide some estimates of the 
kinematics as observed with HST.
\keywords{galaxies: M~31 - 
          galaxies: nuclei --
          galaxies: kinematic and dynamics}
\end{abstract}

\section{Introduction}

Each new observational result concerning the
nucleus of M~31 seems to deepen the mystery of its structure.
This is mainly due to the improvement in the achieved spatial resolution
of the photometric and spectroscopic data: the better resolved 
the object is, the more complex it appears.
For any proposed physical mechanism which can quantitatively
reproduce the observables, it is
important to provide some specific predictions.

In this respect, the nucleus of M~31 is certainly an excellent laboratory
to test our knowledge of galactic nuclei, as we can now optically resolve
scales as small as 0.3 pc (e.g. HST or optically adaptive ground-based 
systems). Indeed, the numerous studies achieved on this object revealed a 
puzzling complexity.

M~31's nucleus is a compact stellar system 
with an average ellipticity of $\ave{\epsilon} \sim 0.4$. 
Asymmetries in its major-axis surface brightness profile (in the visible) 
were already observed in 1974 by Light et al., and subsequently
understood using the pre COSTAR HST data as a double component nucleus
(Lauer et al. 1993). In the V band, the fainter peak (P2) is almost coincident 
with the centre of the bulge isophotes, as the brightest one (P1) is located at
about $0\farcs5$ ($\sim 1.8$ pc for a distance to M~31 of 0.77 Mpc) 
from P2 roughly along the major-axis. At 175 nm however, P2 is the brightest
point, with P1 having an UV upturn similar to its surroundings 
(King et al. 1995).

In 1960, Lallemand et al. obtained spectrograms of the central region
of M~31 and detected the rapid rotation of its nucleus, concluding that
it is a dynamically independent structure. This was confirmed by Kormendy 
(1988)
and Dressler \& Richstone (1988) by long-slit spectroscopy with CCD detectors.
Both studies suggest the presence of a central dark mass of the order of
$10^7$ M$_{\odot}$ in the centre of the nucleus which would account 
for the high value of the central velocity and velocity dispersion gradients.

Sub-arcsecond velocity and dispersion maps were obtained with the
TIGER spectrograph by Bacon et al. (1994, hereafter BEMN94) which uncovered 
more asymmetries (already traced in previous published data):
the velocity field is nearly symmetric about a point $V_0$ located 
very close to P2 ($0\farcs05$ according to Lauer et al. 1993), 
but the maximum $S$ of the stellar velocity dispersion
roughly corresponds to the symmetric point of P1 with respect to P2 
(see Fig.~26 of BEMN94).

Different interpretations have been proposed 
to explain the observed photometric and dynamical asymmetries
in the nucleus of M~31 (see Sect.~\ref{sec:meca} and references therein), 
but they each encounter severe problems. 
Some questions remain open, 
such as the actual efficiency of dynamical friction for an external body
falling into the nucleus,  the life-time of $m=1$ perturbations, 
or the amount of dynamical perturbation on the nucleus itself; we try
here to answer these questions through N body simulations.
In this paper, we concentrate on one of the proposed mechanism, namely
that P1 corresponds to a stellar cluster falling into the potential
well of the nucleus, hosting a massive black hole. 
In a companion paper, we will examine whether a central black hole is
unavoidable, or the presence of a nuclear bar could account for
the observational data.

We summarize and
comment the alternatives already discussed by different authors in Sect.~2.
In Sect.~3 we describe the N body code we used as well as the 
initial conditions of the different experiments. The results are 
discussed in Sect.~4. In Sect.~5, we present and discuss our attempts to
fit the observables using different assumptions.
The general discussion and conclusions are given in Sect.~6.
\section{Different mechanisms}
\label{sec:meca}
The M31 nucleus is an object of peculiar attention, because of
its large central velocity gradient and velocity dispersion: it is
one of the best candidate among nearby galaxies to host a massive black hole 
(BH
of the order of 10$^7$-10$^8$ M$_\odot$, e.g. Kormendy \& Richstone 1995). 
Possible scenarii include one
massive BH, two BHs, or no BH at all. We will consider here only
the first class of models, with several possibilities to account for the
observed asymmetries. Previous studies already discarded a gravitational lens
or a dust component to be responsible for these asymmetries (e.g. BEMN94), 
so we will discuss the two remaining ones:
a dense stellar cluster falling into 
the nucleus (already suggested by Dressler \& Richstone 1988, see also
Lauer et al. 1993, BEMN94) or a thick eccentric disc of stars 
on Keplerian orbits around the black hole, maintained through 
self-gravity and dynamical friction on the bulge (Tremaine 1995).

\subsection{Is the disc eccentric ?}

Tremaine (1995) recently proposed a model based on a set of aligned
Keplerian orbits, forming an $m=1$ pattern. 
The main argument for an orbit crowding effect is the lack of any
significant colour gradient between P1 and the rest of the nucleus,
except for P2 which has a higher UV upturn which could correspond 
to the radio source detected by Crane et al. (1992) and linked with the 
presumed central dark mass. But as argued by King et al. (1995), this 
difference in the UV flux is equivalent to a single PAGB star, although it 
seems more
extended (P. Crane, priv. comm.).
Tremaine (1995) also argued that the observed isophote twist of P1 
as well as the elongations of P1 and P2 along the P1-P2 axis are
natural consequences of the eccentric disc hypothesis. 
The only problem in Tremaine's model lies in the strong predicted 
asymmetry of the observed velocity field, in contradiction 
with the recently acquired
data of Kormendy (SIS/CFHT, $\sigma_{\star} \sim 0\farcs27$).

In his proposition, Tremaine (1995) suggests that the alignment of apsides
of the orbits forming the eccentric disc is maintained by the self-gravity
of the nuclear disc. In fact, an $m=1$ mode could amplify in the disc,
if it has a positive pattern speed $\Omega_p$, and if dynamical friction
against the non-rotating bulge removes effectively the angular momentum, 
provoquing more eccentricity in the disc orbits. 
The efficiency of this mechanism is still unknown.

\subsection{An $m=1$ density wave}
\label{sec:m1}

\begin{figure}
\psfig{figure=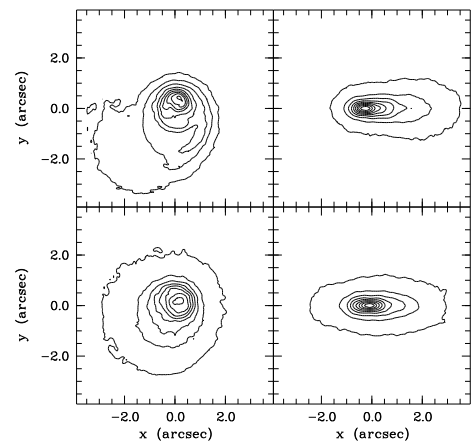,height=6.cm}
\caption[]{Projected isophotes of the nuclear disc with the
black hole launched with an initial offset. Left:face-on, Right:edge-on;
Top: 0.08 Myr; Bottom: 0.32 Myr. }
\label{fig:m1}
\end{figure}

Another possibility is that the $m=1$ perturbation is transient, but
slowly damped. Weinberg (1994) shows that a stellar system can
sustain weakly damped $m=1$ mode for hundreds of crossing times. A fly-by
encounter could excite such a mode, and explain off-centring in some spiral
galaxies. We have tried to estimate the relevant time-scale for the fading
of an $m=1$ excitation in the particular conditions of the M31 nucleus,
in the presence of a massive black hole. The excitation was provided by
a slight initial displacement of the black hole, either in space or
velocity. An $m=1$ spiral wave was generated in the nuclear disc (see
Fig.~\ref{fig:m1}), which disappeared through phase mixing in about 0.5 Myr, 
while the black hole was braked down to the centre by dynamical friction.
The perturbation kept a positive pattern speed $\Omega$= 50km/s/pc
during about 3-4 rotations (the corresponding period is 0.125 Myr).
After that only a material arm was precessing at the negative velocity
$\Omega$= -3.6km/s/kpc until 1 Myr. 
Finally, an unstable $m=1$ mode may also exist in the
presence of gas in the centre of M31 (Shu et al, 1990, Junqueira \& Combes 
1996). 

\subsection{A falling stellar cluster}

Even at HST resolution, the nucleus of M~31 appears as the superposition
of two rather smooth components corresponding to P1 and P2. The detailed
structure of each individual components depends on the decomposition
method (BEMN94, King et al. 1995) but the corresponding 
global parameters seem to vary very little (see Table~\ref{tab:param}).
Let us emphasize, as already did King et al. (1995), that although 
the brightest point of the nuclear region of M~31
corresponds to P1, it never contributes more than 0.55 of the
total surface brightness (assuming P2 is symmetric) and the integrated
magnitude of P1 is 2.6 magnitude fainter than of P2.
\begin{table}
\caption[]{Global parameters for P1 and P2: absolute magnitude in the $V$ band,
core radius, mean ellipticity, relative position angle (PA in degrees),
and displacement (in arcsecond) along the nucleus major ($\delta x'$) and
minor-axis ($\delta y'$).}
\begin{center}
\begin{tabular}{|l|rrrrrr|}
\hline
\# & $M_V$ & $r_c$ & $\ave{\epsilon}$ & $\ave{\mbox{PA}}$ & 
$\delta x'$ & $\delta y'$ \\
 & (mag) & ($\arcsec$)  & & ($\degr$) & ($\arcsec$) & ($\arcsec$) \\
\hline
P1 & -9.6 & $0.25$ & 0.31 & 104 & -0.5 & 0.14 \\
P2 & -12.2 & $0.88$ & 0.39 & 54 & 0 & 0 \\
\hline
\hline
\end{tabular}
\end{center}
\label{tab:param}
\end{table}
Occam's razor suggests to favor the hypothesis of P2 being a nearly
symmetric system, and P1 a superimposed elongated cluster.
Since the velocity profile does not seem to be highly perturbed by the
presence of P1 (even at the resolution of $\sigma_{\star} \sim 0\farcs27$
attained by SIS data), it suggests that the line-of-sight velocity of P1
is close to the line-of-sight velocity of the surrounding nuclear stars
(including the seeing effect). In this scenario, the main effect
is due to the low dispersion of the superimposed cluster which 
produces the offset observed in the velocity dispersion
field (BEMN94). We examine this point in details
in Sects~\ref{sec:disr} and \ref{sec:best}.

\section{N-body models}

\subsection{The code}

We have simulated the infalling of a compact stellar system
in the nucleus of M~31 using an N-body code whose main
characteristics were described in Combes et al. (1990).
A 3D cartesian grid was used, with free boundary conditions, and
we used the method described by James (1977) to avoid a $2^3$ multiplication
of the number of cells in the grid. The total number of particles
was 152384, and the grid size $128\times128\times80$. The time
step of integration was 200 yr.

\subsection{Initial conditions}

As already suggested by Lallemand (1960), the nucleus of M~31 
appears as a distinct photometric and
dynamical component. Indeed, observed properties of the galaxy
exhibit strong changes as we reach a radius of about 3 arcseconds
along the major-axis:
\begin{itemize}
\item the surface brightness rises by more than 1.5 magnitude
in the central 2 arcseconds, while the outer major-axis profile of the bulge
only shows an increase of $0.1$ magnitude in the same interval.
\item the stellar velocity is nearly zero at 3 arcseconds
but peaks around 155 \kms at $0\farcs8$ ($\sigma_{\star} \sim 0\farcs27$,
Kormendy \& Richstone 1995, hereafter KR95).
\item the stellar velocity dispersion stays nearly constant in the inner
bulge with $\sigma \sim 150$ \kms, and starts to rise in the inner
2 arcseconds to reach a value of $\sim 250$ \kms (KR95) in the centre.
\end{itemize}

Moreover, Bacon et al. (1994) have shown that the assumption of the nucleus 
being a dynamically isolated entity leads to differences of less than 1\% in 
the
derivation of a local tensor virial theorem.
This encouraged us to treat the bulge and the nucleus as two separate
components. We therefore adapted the grid to the size of the
nucleus\footnote{Throughout this paper, we use a distance of 0.77~Mpc
for M~31 which leads to a scale of $1\arcsec \sim 3.73$~pc.}, 
by taking a cell size of 0.165~pc which led to a volume of about
$21\times21\times13$~pc$^3$ centred on P2.

\subsubsection{The bulge}

Since the nucleus appears as a dynamically isolated component,
we include the contribution of the bulge as 
a fixed potential\footnote{Note that the definition of the 
``bulge'' depends on the decomposition method. We rely here on the
bulge/nucleus decomposition given in BEMN94.}.

\begin{figure}
\psfig{figure=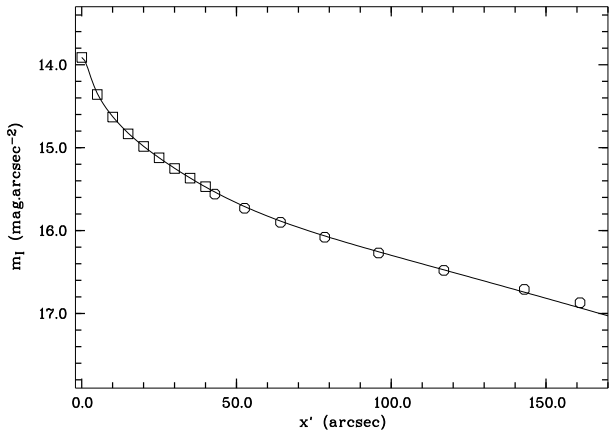,height=6cm}
\caption[]{Major-axis surface brightness profile in the $I$ band
of the fixed bulge compared to the inner axisymmetric bulge model of 
Bacon et al. (1994, squares), and the values given
by Kent outside $40\arcsec$ (circles, rescaled for the $I$ band).}
\label{fig:bulge}
\end{figure}

We have assumed the bulge to be spherically symmetric in our 
simulations. This assumption leads to an error of less than 10\% 
in the potential of the bulge inside the volume of the grid.
Moreover, in the presence of a supermassive black hole of 
a few $10^7$~\Msun, this error is negligible in the central
part of the nucleus.
We thus approximated the mass density distribution of the bulge by a sum
of Plummer spheres of different masses and scales: these functions
yield simple analytic formula for the corresponding 
gravitational potential and forces. 
\begin{eqnarray}
\rho\left(r\right) & = & \sum_i \left(\frac{3 M_i}{4 \pi b_i^3}\right) \times
\frac{1}{\left(1 + \frac{r^2}{b_i^2} \right)^{-5/2}} \\
\Phi\left(r\right) & = & - \sum_i \frac{G M_i}{\sqrt{r^2 + b_i^2}}
\end{eqnarray}
where $M_i$ and $b_i$ are the total mass and the scale of the 
$i^{\mbox{\tiny th}}$ component respectively. We have fitted
the luminosity profile of the bulge using the axisymmetric model of BEMN94
(see Table~11) and the data of Kent (1983). 
This was achieved with a set of 4 Plummer spheres
whose parameters are given in Table \ref{tab:bulge}. Fig~\ref{fig:bulge} 
shows the excellent agreement between this model and the published data
on the inner bulge. 
\begin{table}
\caption[]{Parameters of the Plummer spheres for the bulge component:
total mass ($M_i$), Plummer scale ($b_i$) and the integrated mass
inside a sphere of 10~pc ($M_i^{10}$).}
\begin{center}
\begin{tabular}{|l|lll|}
\hline
$i$ & $M_i$ & $b_i$ & $M_i^{10}$ \\
 & (\Msun) & (pc) & (\Msun) \\
\hline
1 & 3.7 10$^7$ & 16.5 & 5.2 10$^3$ \\
2 & 3.0 10$^8$ & 55 & 1.7 10$^3$ \\
3 & 3.3 10$^9$ & 176 & 6.0 10$^2$ \\
4 & 3.7 10$^{10}$ & 770  & 8.1 10$^1$ \\
\hline
\hline
\end{tabular}
\end{center}
\label{tab:bulge}
\end{table}
The fact that we include the bulge contribution through a fixed
potential allowed us to use all 152384 particles for the nucleus and 
the falling stellar cluster. Following BEMN94, we fixed the total visible 
mass of both components to $1.9~10^7$ M$_{\odot}$, 
leading to a mass of $\sim 125$~\Msun per particle.
The mass of the bulge inside the grid can easily be derived with
the formula: $M(r) = \sum_i M_i r^3 / (r^2 + b_i^2)^{3/2}$.
Inside $r = 10$~pc the bulge contributes for a mass of $\sim 7.6\;10^3$~\Msun,
which is negligible compared to the added masses of P1 and P2.

\subsubsection{The nucleus}

The nucleus is represented by 138530 particles ($\sim 91$\% of 
the total number), with an equivalent mass of $\sim 1.73~10^7$ M$_{\odot}$. 
We simulated it as a Toomre disc of size 3.8~pc. The vertical structure
follows initially a sech$^2$ distribution with a vertical scale of 1.7~pc
at the origin, and slightly increasing with radius to fit the
observed photometry. The radial variation of height with radius was chosen as
$(1+(r/9pc)^2)^{0.35}$.
 The radial velocity dispersion is taken as an exponential law,
to better approach the data:
$$
\sigma_r = \sigma_0 exp(-r/r_v)
$$
with $\sigma_0$=280km/s, and $r_v$ = 5pc. The azimuthal dispersion
is taken, following the epicyclic approximation,
 as $\sigma_{\theta}= \sigma_r {{\kappa}\over {2 \Omega}}$.
 Since the potential is nearly spherical towards the centre due to
the presence of the massive black hole that dominates the mass,
a first approximation of the $z$ dispersion is to take it equal to the
radial dispersion  $\sigma_z=\sigma_r$. However, since this choice was
not enough to keep, after relaxation,  a nucleus as thick as observed, 
 our final choice for the nearly edge-on
nuclear disc was $\sigma_z= 1.3 \sigma_r$.
The actual rotational velocity was computed, after subtracting the
asymmetric drift, according to Jeans equations (cf Binney \& Tremaine 1987). 

\subsubsection{The stellar cluster}

The stellar cluster is represented with the remaining 13854 particles
(9\% of the particles corresponding to $\sim 1.73~10^6$ M$_{\odot}$), 
as a truncated Plummer sphere with an equivalent core radius $r_c = 0.13$~pc
(initial $r_c$ to anticipate any evolution).
The tidal radius $R_t$ of the stellar cluster is adjusted to
its initial environment, since it is launched at a radius $R_0$=6-9pc
from the centre ($R_t$ is between 1.6 and 2.2pc). 
The resulting concentrations $c = \log(R_t/r_c)$
 $\approx 1-1.2$, fall within the whole range of
concentrations observed for globular clusters in our Galaxy (0.75 to 1.75).
The velocities follow an isotropic maxwellian distribution truncated at
the escape velocity, and with a temperature parameter, providing the 
amount of kinetic energy corresponding to the virial theorem.

The cluster is launched in the equatorial plane (at the radius $R_0$=6-9pc) 
with
the velocity $V_0$ (see Table~\ref{tab:init}). In most of the experiments, 
$V_0$ corresponds to the circular velocity at $R_0$. This was motivated by
two facts: first that dynamical friction does not depend much on the 
orbit history of the satellite (e.g. Lin \& Tremaine 1983) provided it is not
yet disrupted. Since the central density of the cluster is an order of
magnitude higher than 10$^5$ M$_\odot$ pc$^{-3}$,
the average density within 6pc including the black hole,
we know that the tidal disruption has not occured at this radius
(only the envelope of the cluster has been stripped away, which is taken
into account through the initial $R_t$).
And second, dynamical friction has the tendency
to circularise orbits (Bontekoe \& van Albada 1987).
However, we have in Sect.~\ref{sec:disr} also performed experiments with 
a cluster on an hyperbolic orbit: this will give us insights for the 
particular case of M31's nucleus (see Sect.~\ref{sec:best}).

\subsubsection{The black hole}

The observed steep central velocity gradient and velocity dispersion peak
suggested the presence of a central dark mass of a few $10^7$~\Msun
(Dressler \& Richstone 1988, Kormendy 1988). Bacon et al. (1994)
confirmed this result with self-gravitating two-integral dynamical models of 
the nucleus of M~31, and derived a central dark mass of $\sim 7.10^7$~\Msun.
We model it with a Plummer sphere with a core radius $r_c = 0.13$~pc 
(i.e. limited by our resolution of 0.15 pc) and a mass of a few times
$10^7$~\Msun (see Table~\ref{tab:init}).

\subsection{Initial parameters}
 We restrict our analysis to 
some of the parameters we thought to be crucial for the
evolution of the system (cf Table~\ref{tab:init}):
\begin{itemize}
\item the black hole mass $M_{bh}$,
\item the mass-to-light ratio of the cluster $(M/L)_C$ (i.e. its mass),
\item the initial radius of the cluster $R_0$.
\item the initial velocity of the cluster $V_0$.
\end{itemize}

\begin{table}
\caption{Parameters of the different experiments: the black hole mass $M_{bh}$
(\Msun), the ratio between the $M/L$ of the nucleus and the cluster 
$\eta(M/L)$,
the core radius of the cluster $r_c$ (pc), the radius $R_0$ (pc) and velocity
$V_{0}$ (\kms) at which the cluster is launched.}
\begin{center}
\begin{tabular}{|l|r|rrrr|}
\hline
run & $M_{bh}$ & $\eta(M/L)$ & $r_c$ & $R_0$ & $V_{0}$ \\
\hline
\#1 &  $6.5\;10^7$ & 1 & 0.12 & 8.8 & $V_c$ \\
\#2 &  $9.0\;10^7$ & 1 & 0.12 & 6.6 & $V_c$ \\
\#3 &  $1.1\;10^8$ & 1 & 0.12 & 6.6 & $V_c$ \\
\#4 &  $1.1\;10^8$ & 2 & 0.12 & 6.6 & $V_c$ \\
\#5 &  $1.1\;10^8$ & 1 & 0.12 & 6.6 & $1.5 \times V_c$ \\
\#6 &  $1.1\;10^8$ & 1 & 0.12 & 6.6 & $V_c$ \\
\#7 &  $1.1\;10^8$ & 1 & 0.12 & 6.6 & $- V_c$ \\
\#8 &  $1.1\;10^8$ & 1 & 0.12 & 6.6 & $V_z$ \\
\#9 &  $2.2\;10^7$ & 1 & 0.12 & 6.6 & $V_c$ \\
\hline
\hline
\end{tabular}
\end{center}
\label{tab:init}
\end{table}
\section{The disruption of the cluster}
\label{sec:disr}

If P1 indeed corresponds to an additional
cold stellar system and is spatially close
to the centre of the nucleus, its decaying time 
should be rather short, and the probability 
to observe such a configuration is very small.
This is the main argument against the falling cluster hypothesis.
However Lauer et al (1993) emphasize that the dynamical friction 
effects are not well known, and there exist some circumstances where
it may shut off. They privileged the hypothesis of P1 being an
external stellar system coming from another nucleus, possibly having
its own central BH.

\subsection{Dynamical friction}

A simple application of the Chandrasekhar formula
(see e.g. Binney \& Tremaine 1987)
can give us a first estimate of the falling time of a stellar cluster.
For a system of 1.6~$10^6$~\Msun, it would take a Hubble time to
fall in the centre from a radius of a few kpc. 
This timescale is strongly dependent on the initial radius $r_i$, so that only
$\sim 10^5$~yr are needed for the system to fall from a radius of 6.6~pc
(with $V_{P1} \sim 300$~\kms, see Table~\ref{tab:init} and 
Sect.~\ref{sec:geom}). 

However, the application of the Chandrasekhar formula here is
not appropriate. On one hand,
this formula overestimates the dynamical friction since it assumes
a singular isothermal sphere while P1 has entered the core where the density
remains constant. Moreover, the gravitational potential is dominated by a 
central point-like mass, which does not provide friction.

On the other hand,
our simulations confirm a much more important result already
noticed in the case of interacting galaxies by Prugniel \& Combes (1992):
the Chandrasekhar formula underestimates the effect of dynamical friction 
since it assumes that the accreted system is rigid. Prugniel \& Combes (1992) 
showed that releasing the hypothesis of a rigid satellite increases 
considerably the friction efficiency. In our experiments, the main dynamical 
friction is indeed due to the deformation of the stellar cluster, 
a reciprocal effect usually neglected in estimations based on the 
Chandrasekhar formula.

\subsection{Tidal forces and the black hole mass}

\begin{figure}
\psfig{figure=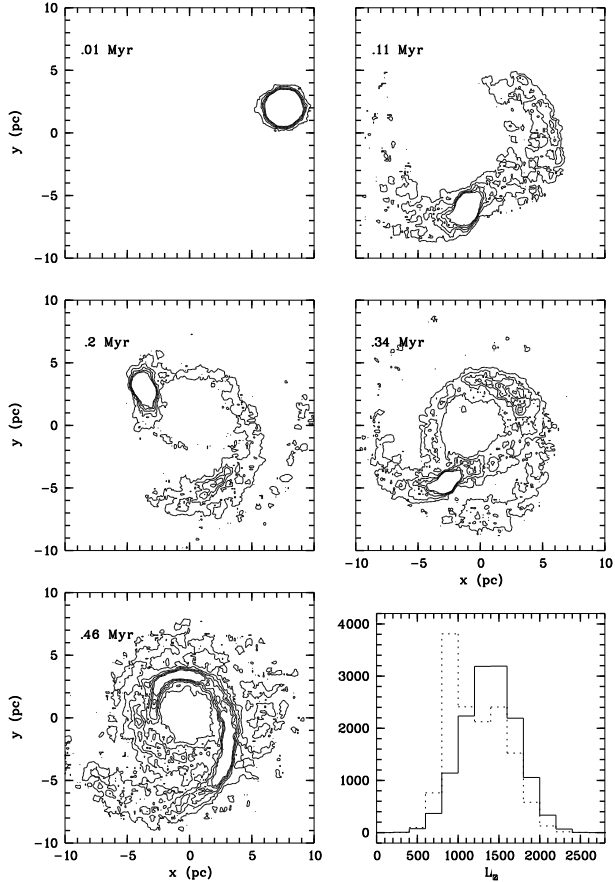,height=12cm}
\caption[]{Face-on view of the cluster particles at different times
of the simulation (run \#1): the surface brightness has been
slightly smoothed and only the faintest contours have been 
drawn to emphasize the development of the ring-like structure. 
The bottom right panel shows the histogram of the angular momentum
for the cluster particles at the begining and at the end of the
simulations (0.01 Myr -- solid line -- and 0.46 Myr -- dashed line --
respectively).}
\label{fig:ring}
\end{figure}

\begin{figure}
\psfig{figure=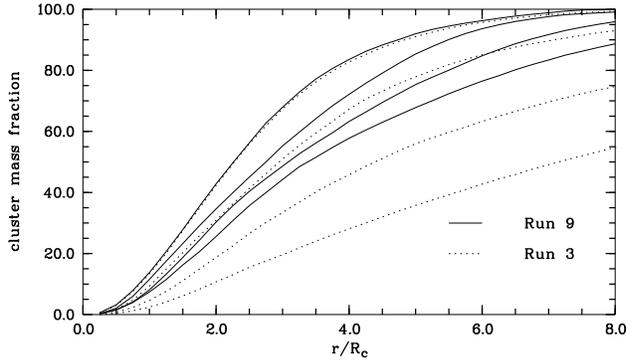,height=5cm}
\caption[]{Effect of the central mass concentration on the disruption 
of the cluster: the mass of the cluster (in \% of its total mass)
enclosed in a sphere of radius $r$ (normalized by the initial core radius
$r_c$) is shown as the simulation evolves
(from top to bottom, $t=\;10^4$, $2.10^4$, $3.10^4$ and $4.10^4$),
for two different central mass concentrations
($2.10^7$~\Msun -- solid lines -- and $10^8$~\Msun -- dotted lines).}
\label{fig:bhmass}
\end{figure}

Our simulations show that the main effect is due to
the tidal forces induced by the central mass concentration which rapidly 
disrupt the cluster. The strength of the tidal forces strongly depends
on the ratio of the central cluster and nucleus concentrations.
For a black hole mass of $\sim 6.10^7$~\Msun, the cluster looses 
30\% of its mass in less than $10^5$ yr. After $4.10^5$ yr, they are all 
spread in a ring-like structure (Fig.~\ref{fig:ring}),
{\em although the centre of gravity of the cluster's 
particles is close to the centre of the nucleus}.
For masses greater than 6~$10^7$~\Msun, the black hole dominates
the overall gravitational potential in the central 10~pc. 
The lifetime of the cluster is then roughly scaled like the inverse of the 
square root of the central dark mass (Fig.~\ref{fig:bhmass}).
We also ran an experiment equivalent to run~\#3 but without the 
additional central mass.  
In this run, the stellar cluster is not disrupted. On the contrary, it is the 
nucleus now that suffers strong tidal deformations, and the cluster
decays like a rigid body through dynamical friction, produced by the
nucleus deformations. The decay time is then longer, of the order
of 3 10$^5$ yr.

In Fig.~\ref{fig:ring}, we can also see that there is very little transfer of 
angular momentum between the nucleus and the original cluster:
the mean angular momentum had only changed by 10\% by the end of the simulation
at 4.6~$10^5$ yr. 
This result is very similar to the one obtained by Charlton \& Laguna (1995)
at a larger scale: the debris of the globular clusters disrupted by tidal 
forces mainly follow the original cluster orbit (i.e. highly eccentric orbits).

\begin{figure}
\psfig{figure=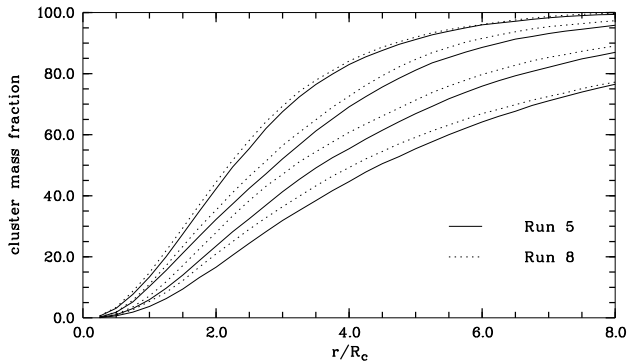,height=5cm}
\caption[]{Same as Fig.~\ref{fig:bhmass} but for two different 
initial velocities of the cluster:
run \#5 corresponds to a cluster on an hyperbolic orbit in the equatorial 
plane,
and run \#8 to an initial pure vertical velocity for the cluster.
Note the difference with run \#3 in Fig.~\ref{fig:bhmass}.}
\label{fig:velo}
\end{figure}

In most runs presented in this paper, we have launched the cluster with the 
circular velocity. In run \#5, we have launched the cluster
with a tangential velocity of $1.5\times V_c$ (hyperbolic orbit)
and in run \#8 with a purely vertical velocity of 250 \kms. 
In both cases, the distance between the cluster and the central
mass concentration increases with time.
This prevents the cluster to be strongly disrupted as illustrated in
Fig.~\ref{fig:velo} (see also Charlton \& Laguna 1995).

\subsection{The mass-to-light ratio of the cluster}

Because of the absence of colour gradients, we
assumed until now that P1 and P2 have similar stellar
populations and mass-to-light ratios. We have however
also explored different mass-to-light ratios
(hereafter $M/L$).
For the run \#4, we fixed the $M/L$ of the cluster to be half
the one of the nucleus (see Table~\ref{tab:init}).
This was achieved by assigning fewer particles to the cluster,
which were then used for the nucleus whose total mass was conserved. 
The less massive cluster is obviously less concentrated at all times
after the launch. However, its envelope seems to survive longer.
To understand this point it is necessary to remember that the initial
total energy of the clusters are different in those two runs:
in run~\#3 the cluster covers a longer track which induces a 
larger spreading of the unbound particles. Moreover, 
the tidal radius of the cluster is significantly smaller 
(by about 25\%) in run~\#4 than in run~\#3.
\section{The case of M31 nucleus}
\label{sec:best}

\subsection{Some simple geometrical arguments}
\label{sec:geom}

Let us first assume that the stellar cluster is located 
in the equatorial plane of the (axisymmetric) nucleus, and let $(x^c, y^c)$,
$(V_x^c, V_y^c)$ be its position and velocity in a reference system centred on
P2 where $Ox$ is the line of nodes.
Since we observe P1 at about $0\farcs14$ of P2
along the nuclear minor-axis and $0\farcs5$ along
the major axis  (Lauer et al. 1993, BEMN94), 
we can write: $y^c = 0\farcs14 / \cos{i}$ with $i$ being the 
inclination angle of the nucleus ($90\degr \equiv$ edge-on). 
In this case, the nucleus of M~31 cannot be viewed exactly edge-on.
The average axis ratio of the nucleus of $\ave{q}$ of 0.61
can be used to bracket its inclination angle: 
$\arccos{\left(0.61\right)} \sim 52.4 < i (\degr) < 90$.
Then, the apparent distance of P1 to the major-axis of the nucleus
provides us a lower limit for the true distance of 
P1 to the line of nodes: $y^c > 0\farcs14 / 0.61 \sim 0\farcs23$.
And we have now $R^c = \sqrt{({x^c}^2 + {y^c}^2)} > 0\farcs55$
($\sim 2$~pc at 0.77 Mpc). If the near side of the nucleus corresponds
to the near side of the main disc of M~31, i.e. the North-West
side  (see Fig.~3 of Kormendy 1988),
we can conclude from the sign of the observed velocities
that P1 is going toward the line of nodes of the nucleus.

The projection on the sky gives $V_y^c = V^c_{los} / \sin{i}$,
where $V^c_{los}$ is the line-of-sight velocity of P1.
If we assume now that the cluster is on a circular orbit, we can also write:
\begin{equation}
V^c \left( R \right) = \frac{V^c_y \times R^c}{x^c}
\label{eq:vp1}
\end{equation}
with $V^c$ the spatial velocity of P1. Using Fig.~7 of Kormendy
\& Richstone (data of Kormendy \& Bender 1995), we estimate 
$V^c_{los}$ to be $\sim 160$~\kms (see Sect.~\ref{sec:best}).
This value is very uncertain, but the following calculations
can be adapted accordingly.
The projected distance $x'$ ($= x$) along the nuclear
major-axis is $\sim 0\farcs5$ (Lauer et al. 1993, BEMN94). Therefore 
$V^c \left( R \right) \sim 320 \cdot (\sqrt{(0.25 + {y^c}^2)} / \sin{i})$.
$V^c \left( R \right)$ is then minimum ($V^c \sim 211$~\kms) for
$i \sim 62.6\degr$ which gives $y^c \sim 0\farcs3$ and $R^c \sim 0\farcs59$.

\begin{figure}
\psfig{figure=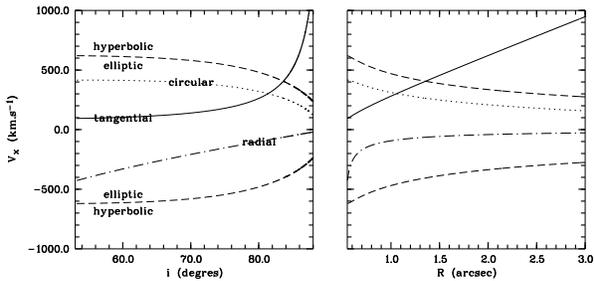,height=4cm}
\caption[]{Diagrams showing the different types of orbits
for a test particle in the modelled potential of M~31 (see text)
in terms of $i$ (or $R^c$) and $V_x^c$: the dashed lines correspond to the
escape velocity, the dotted line to the circular velocity.
Geometrical constraints are given by the solid line (pure
tangential velocity) and the dotted--dashed line (radial orbits).
Vertical dotted lines indicate the circular velocity solution.}
\label{fig:vp1}
\end{figure}

In fact, since we already have observed values for $x^c$ and $V_{los}^c$, 
we only need to get $y^c$ (or the inclination $i$) and $V_x^c$
to know the full position and velocity of the cluster.
Using the model given in Bacon et al. (1994) for the spatial
mass distribution and assuming a supermassive black hole of $M_{bh} 
= 10^8$~\Msun (see Sect.~\ref{sec:best}), it is then possible
to roughly constrain the type of orbit of the cluster
in terms of these two unknowns $y^c$ and $V_x^c$
(as a test particle in a fixed potential).
This is illustrated in Fig.~\ref{fig:vp1}, where the limits
of the regions for elliptic, and hyperbolic orbits are drawn as well
as the curves for tangential and radial velocities.
We thus find that there is only one point where P1 
can be on a circular orbit: $V^c \sim 300$ \kms for $R^c 
\sim 1.05\arcsec$ and $y^c \sim 0\farcs9$ (see Fig.~\ref{fig:vp1}). 
Taking into account the uncertainty on the true
value of $V^c_{los}$ and on the circular velocity in the nucleus,
Fig.~\ref{fig:vp1} indicates that $0.85 < R^c (\arcsec) < 1.3$
or $0.7 < y^c (\arcsec) < 1.2$ ($2.6 < y^c (pc) < 4.5$ at 0.77 Mpc).

The core of the cluster could survive
only if its central density is larger than the average mass
density of the nucleus interior to its orbit.
We have seen in the simulations that the initial central density
of the cluster had to be $\approx$ 10 times the average density at
$R_0$ = 6pc (including a black hole mass of 7 10$^7$ M$_\odot$).
We therefore expect that the disruption will occur
at a radius of $\sim 3$~pc. This roughly corresponds
to the radius derived for a circular orbit inside the nucleus.
If the stellar cluster follows such an orbit it is then presently 
being disrupted. Such quasi circular orbits were tried, where
the cluster was slowly spiraling in towards the centre
because of dynamical friction. But the stripping of its outer parts
was so effective that, although the center of mass of the system
did reach the required radius, the remaining core of the globular
cluster did not. Also, the disruption of the cluster produced
a much too large internal velocity dispersion to the stellar
debris around the core; the presently observed low velocity dispersion 
of P1 strongly suggests that this is not the case yet. Therefore,
we must conclude that P1 cannot be on a circular orbit
around P2.

Finally, for a central dark mass of $1.\;10^7$~\Msun
the circular velocity predicted from the model given in BEMN94
is smaller than 160~\kms at radii larger than $0\farcs55$.
Therefore, if the mass concentration is smaller than
$1.\;10^7$~\Msun, the cluster is on an orbit unbound
with respect to the nucleus, and will escape from
its gravitational influence (and might eventually fall back later
by friction).

We have not explored many different inclinations of the orbital plane.
However, as the main
influence comes from the central dark mass, the fate of the cluster
is mainly determined from its pericentre distance.

\subsection{Observations of the experiments}

The main problem to fit self-consistently the nuclear disc of M31 is
its observed thickness, mainly under the assumption of its nearly
edge-on inclination. This requires a relatively high $z$-dispersion,
while the disc is highly rotating in the plane. 
The system relaxes quickly, on time scales of the order of 10$^4$ yr,
and it revealed very difficult to maintain
both a rather thick disc, and a high V$_{rot}/\sigma$ ratio.
With relaxation, the velocity dispersion tends to become more isotropic, 
and the rotational velocity decreases with respect to the in-plane 
dispersion. To reproduce the high azimutal velocities, a more massive black 
hole is required, and no good fit was obtained below 8 10$^7$ M$_\odot$.

We have ``observed'' the simulations by associating a specific luminosity
of 50.5~\Lsun per particle as to reproduce the photometric data in the
$I$ band (i.e. $M/L_I \sim 2.5$). 
For the bulge, we used the axisymmetric surface brightness
model given in BEMN94 (including the difference in position
angle, see their Table~4).
The system (nucleus + cluster) was viewed such as the strongest projected peak
of the cluster is located at $x' = 1.85$~pc and $y' = 0.5$~pc 
from the centre of the nucleus
(see BEMN94). This fixes the three Euler angles $(\psi, \theta, \phi)$
of the projection. All the relevant characteristics 
of the various instrumental set-ups are then taken into account:
spatial and spectral pixel sizes, spatial and spectral resolutions.
The Point Spread Function was approximated by a single gaussian function
which is thus completely fixed by its width $\sigma_{\star}$.
For the bulge we assumed a gaussian Line Of Sight Velocity Distribution 
(hereafter LOSVD) whose first two moments are those of the corresponding 
isotropic
axisymmetric model (i.e. Model A of BEMN94 with $M/L_I = 2.5$ and the
appropriate central dark mass). The obtained total LOSVDs were parametrized
with the help of Gauss-Hermite functions (see van der Marel \& Franx 1993)

Let us first examine the hypothesis of a projection effect, where
the cluster is physically far from the centre of M~31. 
The results of these experiments will be used to constrain a fully 
self-consistent simulation.

\subsection{Stellar cluster seen in projection against the nucleus}
\label{sec:incl}

Lauer et al. (1993) argued that the observed photometry and kinematics
speaks against a cluster physically far from the centre and seen in projection:
\begin{itemize}
\item P1's luminosity profile is truncated and asymmetric:
this is indeed suggestive of an interaction between P1 and P2.
However, as mentioned by King et al. (1995), the decomposition
is certainly not unique 
\item Lauer et al. (1993) also argued that the positioning of P1
on the nucleus major-axis favors a physical relationship
between P1 and P2. But this assumes that P2 is seen almost edge-on.
Moreover, P1 is closer to the major-axis of the bulge.
\item The velocity profiles are symmetric, as though P1
 did not introduce any perturbation. But this could come 
in part from the fact that the observed kinematics is luminosity 
weighted (seeing effect, cf Sect.~\ref{sec:hstres}).
\item Lauer et al. (1993) estimated that the P1 velocity dispersion
is too large for a globular cluster.
However, the recent SIS data obtained by Kormendy indicate
that $\sigma_{P1} < 85$~\kms (KR95). Moreover, our simulations suggest
(see Sect.~\ref{sec:compkin}) that $30 < \sigma_{P1} ($\kms$)< 80$
(if P1 is being disrupted it is certainly not in virial equilibrium).
\end{itemize}
Although P1 and P2 seem to be interacting,
there are thus no definite arguments against a projection
effect. We will therefore examine this possibility in the
following paragraphs, and use this to test the stability of the
model, after a relaxation of several dynamical times.

The apparent flattening of the nucleus (see Sect.~\ref{sec:geom})
sets a lower limit of $\sim 53\degr$ for its inclination angle.
We have attempted to reproduce the photometry and the kinematics
of M31~'s nucleus using a thin disc and a central dark mass
of a few $10^7$~\Msun. We managed to get a reasonable fit with
$M_{bh} = 7.7\;10^7$~\Msun and an inclination of $55\degr$.
The cluster has been fixed as a separate Plummer sphere with $r_c = 0.8$~pc,
projected onto the nucleus. The simulation of the nucleus has 
been continued until the system reached dynamical stability. 
Self-consistent experiments are described in Sect.~\ref{sec:edge}.

In Fig.~\ref{fig:hst31} we present the isophotes\footnote{In all the following 
figures, we have aligned the nucleus major-axis with the horizontal axis:
see BEMN94.} of the model compared to
the WFPC2 $I$ band image kindly provided by Michael Rich.
We normalized both images using the HRCAM $I_c$ band image (BEMN94).
The fit is rather good, although we do not reproduce the
flattening of P1 as well as its full spatial extension (see King et al. 1995).
Our nucleus is also slightly more peaked than in the observations.
This is due to the fact that we start from conditions that are close
to what is observed. Then dynamical relaxation is at work and the nucleus
has a tendency to concentrate.
\begin{figure}
\psfig{figure=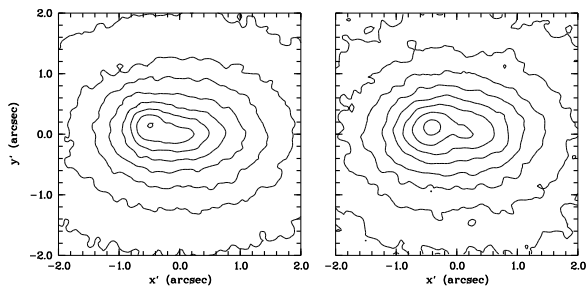,height=4cm}
\caption[]{$I_c$ band isophotes of the ``inclined'' model (left panel)
and of the WFPC2 $I$ band image (normalized to the Cousin system, right panel).
The brightest isophote corresponds to 11.85 \magsec, and the 
step is 0.3 \magsec.}
\label{fig:hst31}
\end{figure}

\subsubsection{Comparison with published kinematics}
\label{sec:compkin}

As discussed above, the line-of-sight velocity of the cluster $V^c_{los}$ 
suggested by the available observables is probably close to 160~\kms.
However, it is difficult to determine $V^c_{los}$ with certainty.
A way to constrain the mean value for $V^c_{los}$ is to directly
look at the influence of P1 on the LOSVDs. This has been already done
by BEMN94 who unfortunately only examined a very limited subset of 
possibilities. Moreover, at the spatial and spectral
resolution of the TIGER data ($\sigma_{\star} \sim 0\farcs38$)
the light contribution of P1 is significantly diluted. BEMN94 still
observed a slight asymmetry in the velocity curve along the nucleus
major-axis which hinted for a rather high apparent value of $V^{P1}_{los}$
($> 120$~\kms). We have fixed $V^c_{los}$ to a number of different
values to test for its effect on the shape of the resulting LOSVDs
and the corresponding derived velocities and dispersions. 
The internal velocity dispersion of the cluster is assumed
isotropic with $\sigma^c \sim 36$~\kms.

\begin{figure}
\psfig{figure=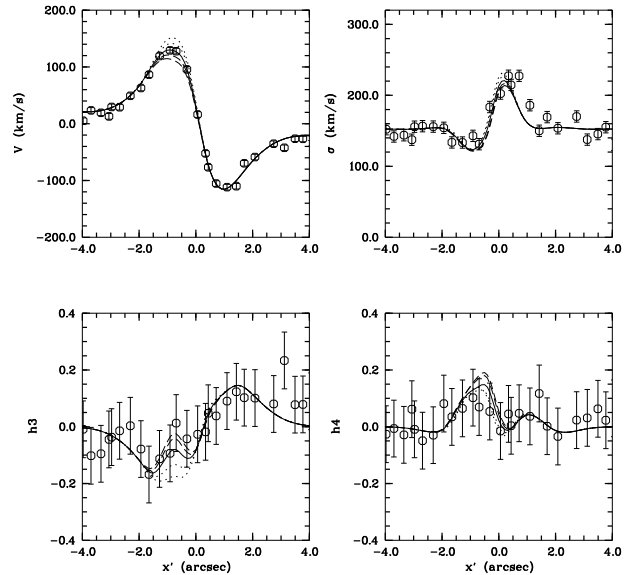,height=8cm}
\caption[]{TIGER velocity, dispersion, $h_3$, and $h_4$ major-axis 
profiles (circles). The corresponding profiles derived from the ``inclined 
model''
are presented for different values of $V^c_{los}$:
100 and 120~\kms (dashed lines), 140~\kms (solid line), 
160 and 180 \kms (dotted lines).}
\label{fig:ptiger31}
\end{figure}

For comparison we have used the kinematics obtained
with the two-dimensional TIGER data presented by BEMN94
(spatial resolution of $\sigma_{\star} \sim 0\farcs37$ and
spectral resolution of $\sim 100$~\kms). 
The best fit is obtained for $V^c_{los} = 140$~\kms
but values of 120 and 160~\kms are also consistent with the data
(Fig.~\ref{fig:ptiger31}). Higher velocities for the cluster tend to increase 
both
the peak in the velocity and velocity dispersion profiles,
but to reduce the influence on higher order moments.
It is important to remember that the kinematics
were derived here from a gaussian fit to the LOSVDs, and that
true statistical moments are less influenced
by the presence of the cluster.
Although Fig.~\ref{fig:ptiger31} only shows the major-axis kinematical
profiles, the agreement is excellent on the full two-dimensional fields.
The higher resolution data of Kormendy (SIS/CFHT, $\sigma_{\star} = 0\farcs27$;
KR95) also suggests a value of $V^c_{los} \sim 160$~\kms,
which will therefore be used for the ``self-consistent'' model 
(Sect.~\ref{sec:edge}).

We have finally estimated the effect of a higher velocity dispersion
of the cluster by increasing its mass by a factor of 2:
this increases the internal velocity dispersion by a factor of $\sqrt{2}$.
As expected the velocity profiles are almost coincident but the velocity
dispersion profiles differ significantly: the minimum along the major-axis
reaches $\sim 95$~\kms for $\sigma^c = 36$~\kms ($M/L_I = 2.5$)
and $\sim 112$~\kms for $\sigma^c = 51$~\kms ($M/L_I = 5$)
closer to the observed values.
This suggests a reasonable interval for the internal dispersion
of the cluster: $30 < \sigma^c ($\kms$) < 80$.

\subsection{Stellar cluster inside the nucleus}
\label{sec:edge}

We have then examined the case of a rather more edge-on
model for the nucleus, but this time including the
stellar cluster in a self-consistent way. In this case, the nucleus
should be significantly thicker as the apparent flattening
is closer to the true one, and we have chosen a $z$ velocity dispersion 
slightly higher than the radial dispersion ($\sigma_z=1.3\sigma_r$,
see Sect.~3.2.2). 
The nucleus is projected with an inclination $\sim 70\degr$. The orbit of
the stellar cluster is no longer in the equatorial plane, but follows a nearly
planar orbit, inclined by $\sim 20\degr$, so that the orbit is almost 
edge-on projected on the sky.
We have already excluded a circular orbit (\ref{sec:geom}), and therefore
have chosen an hyperbolic orbit, the cluster being now very
close to its pericenter of 3pc, where it has
barely reached the critical radius of disruption.

\begin{figure}
\psfig{figure=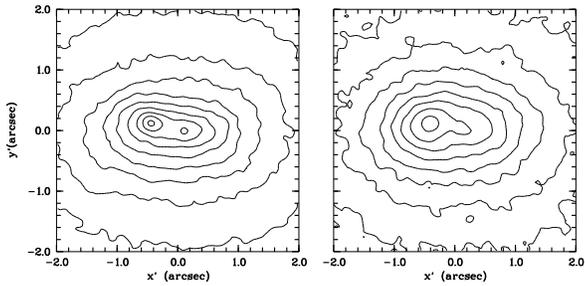,height=4cm}
\caption[]{Same figure as Fig.~\ref{fig:hst31} but for the 
``self-consistent'' model.}
\label{fig:hst64}
\end{figure}

We obtained a reasonable fit to the observables for
a central dark mass of $9.4\;10^7$~\Msun and an inclination
of $68\degr$ (constrained by the observed
position of the cluster). The cluster was launched
at 8.8~pc from the centre with a velocity of 420~\kms. 
The cluster has a strong effect on the stability of the nucleus:
the outer parts gain energy and inflate, while the center becomes
more condensed. This is difficult to anticipate in the initial
conditions, and causes the main discrepancy between this model and the 
observables: the central surface brightness
of P2 is larger in the model than in the HST photometry 
(Fig.~\ref{fig:hst64}). Also,  
the nucleus has the tendency to dynamically relax towards a more
centrally concentrated and flattened system, due to the strong influence
of the central dark mass. The latter completely dominates the
gravitational potential in the inner few arcseconds and tends to
symmetrize the dispersion tensor in the centre.
We should note however that although P2 is too concentrated 
in the model it has the correct integrated magnitude.

\begin{figure}
\psfig{figure=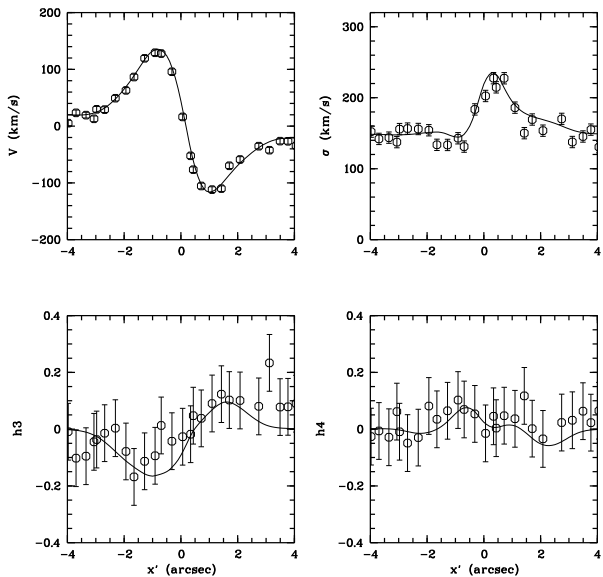,height=8cm}
\caption[]{Same as Fig.~\ref{fig:ptiger31} but for the 
self-consistent'' model.}
\label{fig:ptiger64}
\end{figure}

\begin{figure}
\psfig{figure=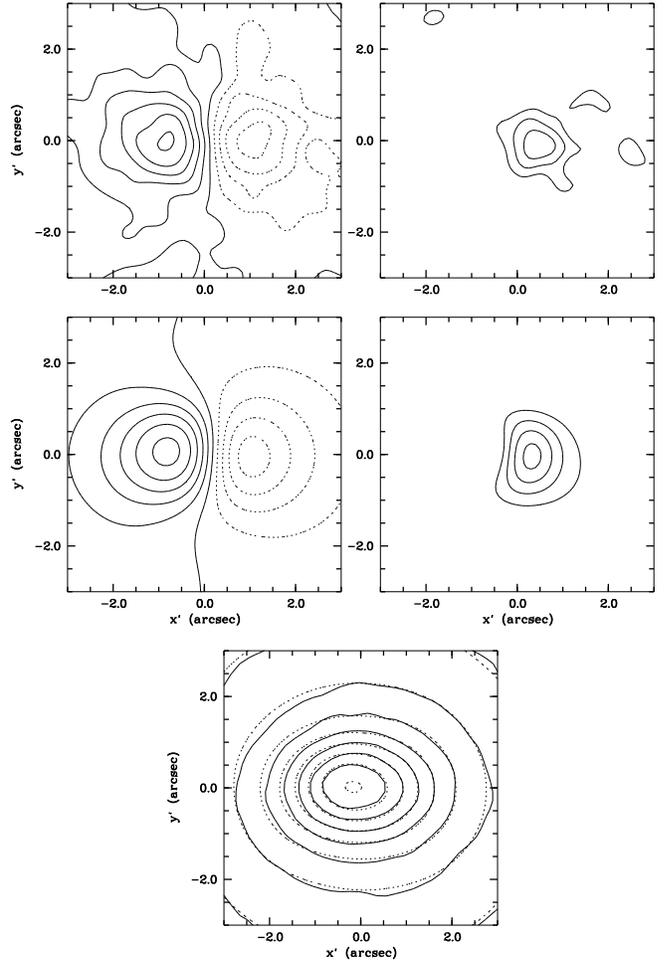,height=13cm}
\caption[]{Isophotes (bottom), velocity fields (left panels)
and velocity dispersion fields (right panels) from the
TIGER data (top panels) and of the ``self-consistent model'' (middle panels). 
The steps are respectively 0.25 magnitude, 25~\kms ($V$)
and 15~\kms ($\sigma$).}
\label{fig:tiger64}
\end{figure}

\begin{figure}
\psfig{figure=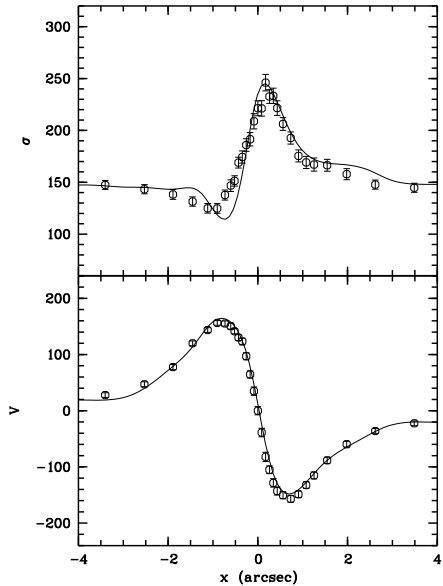,height=8cm}
\caption[]{Velocity and velocity dispersion of the SIS data (circles,
KR95 and of the ``self-consistent  model''.}
\label{fig:korinc2}
\end{figure}

\begin{figure}
\psfig{figure=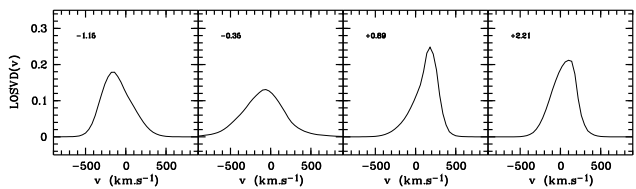,height=2.6cm}
\caption[]{LOSVDs derived from the ``self-consistent model'' at
four different positions to be compared with Fig.~7 of
KR95.}
\label{fig:panel64}
\end{figure}

Figs.~\ref{fig:ptiger64}, \ref{fig:tiger64},
\ref{fig:korinc2} and \ref{fig:panel64}
present the comparison of this model observed at 
$t = 1.6 \; 10^4$~years after the launch of the cluster
with the TIGER and SIS data. 
The agreement is rather good even for the higher order Gauss-Hermite
moments $h_3$ and $h_4$ (Fig.~\ref{fig:ptiger64}).
As mentioned above, the isophotes are slightly too flattened
which leads to a higher central surface brightness than in the
observed image.
At a time $t = 1.6\; 10^4$~years, the cluster which lies at
about $1\farcs1$ (4.1~pc) is already partially disrupted,
although the core still survives with a projected FWHM
of $\sim 0\farcs4$. This is slightly smaller than the 
value of $0\farcs5$ FWHM derived by King et al. (1995). 
The spread of the particles of the cluster induced by the tidal
forces is accompanied by the creation of a large internal velocity gradient
which in projection goes from $\sim -105$~\kms to $\sim 240$~\kms.
This dilutes the contribution of the cluster to the observed LOSVDs.

\subsection{M~31 at the HST resolution}
\label{sec:hstres}

At the HST resolution, the observed kinematics of the nucleus 
of M~31 should clearly reveal the presence of a superimposed
cold cluster. This is illustrated in Fig~\ref{fig:slithst} where
we have derived the kinematics of the ``self-consistent model'' 
assuming a pixel size of $0\farcs06 \times 0\farcs06$.
We also present seven corresponding LOSVDs along the line joining
P1 and P2 for squared apertures\footnote{These parameters were chosen 
to be similar to the characteristics of the observations of M~31 with the
HST Faint Object Spectrograph.}
of $0\farcs26$ and $0\farcs09$ and a spectral resolution of 100~\kms.
Note the large wings of the LOSVD at the central position due to the
presence of the mass concentration. There are no significant differences
between the observed LOSVDs for the two different apertures.
The velocity (gaussian fit) along the major-axis reaches $\sim -220$~\kms
on the side opposite to P1 at $\sim 0\farcs45$, and the maximum dispersion 
is $\sim 290$~\kms at $\sim 0\farcs25$. The maximum positive velocity and
the minimum dispersion obviously strongly depend on the true internal 
kinematics of P1.
\begin{figure}
\psfig{figure=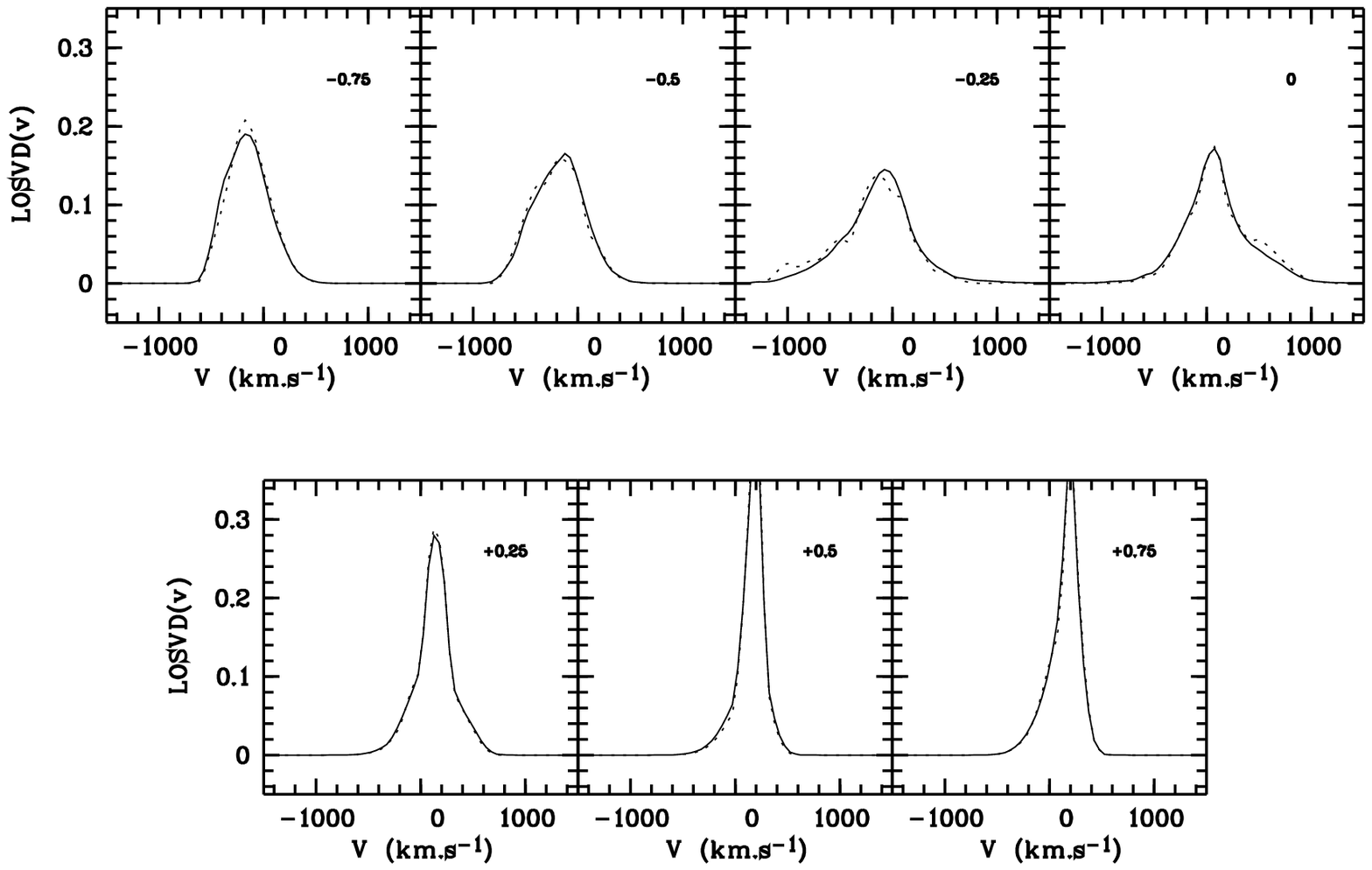,height=5.3cm}
\caption[]{LOSVDs from the ``self-consistent model''
at $-0\farcs75$, $-0\farcs5$, $-0\farcs25$, $0$, $+0\farcs25$,
$-0\farcs5$, $+0\farcs75$ along the P1/P2 line for squared apertures
of $0\farcs26$ (solid lines) and $0\farcs09$ (dotted lines).}
\label{fig:loshst}
\end{figure}
\begin{figure}
\psfig{figure=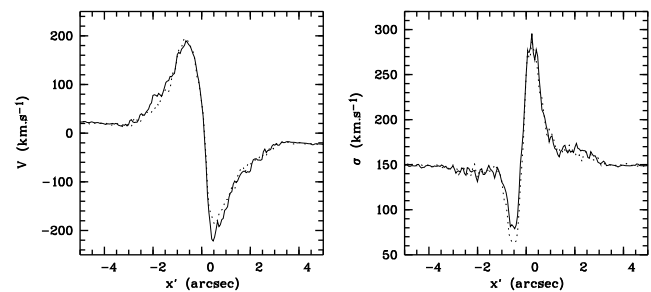,height=4cm}
\caption[]{Velocity (left) and dispersion profiles (right) along the
major-axis of the nucleus (solid lines) and along the P1/P2 line
(dotted lines) predicted from the ``self-consistent model''.}
\label{fig:slithst}
\end{figure}
The main prediction for spectrographic observations at the HST resolution
is the strong asymmetry in the velocity profile along the P1-P2
line, which is also clearly revealed in the shape of the LOSVDs. 
However, whatever P1 corresponds to (a stellar cluster
or a density perturbation), the fact that the observed
kinematics is weighted by light induce strong asymmetries.
Finally, as noticed by Lauer et al. (1993), 
if the cluster has indeed a transverse velocity of $\sim 440$~\kms
as in the ``self-consistent model'' presented here,
its proper motion should be $\sim 10^{-3}$~arcsecond in 10~years.

\section{Discussion and conclusions}
\label{sec:disc}

In this paper we have presented N-body simulations of
a stellar cluster falling towards a galactic nucleus.
These experiments were designed to ressemble the 
central region of M~31. 
First we have shown that it was possible to obtain 
reasonable fits to all observables, either with 
a projected stellar cluster (which served to
adapt our initial conditions for the N body
simulations), or with a stellar cluster close
to the nucleus and interacting with it.
The best model considered an initial orbit
where the cluster was unbound to the nucleus.
Slight discrepancies between this model and the data still remain
(e.g. the nucleus isophotes flattening). This
solution is certainly not unique and other significantly
different initial conditions (e.g. the choice of the orbit)
could lead to very similar observables.

Second, we have shown that the 
dynamical friction is even more efficient than
previously estimated: the cluster rapidly decays to
smaller distances, where it is disrupted through
tidal interaction with the BH. The lifetime
of such a cluster is short ($< 5 \; 10^5$) at radii
smaller than $\sim 10$~pc if a central dark mass
of a few $10^7$~\Msun is present at the centre.
The main argument against the scenario of a falling
cluster is that we must see the M31 nucleus at a very
special time. However, the decaying of a stellar cluster is not
a rare phenomenon.
Tremaine et al (1975) proposed that the falling of globular clusters
could even be the formation mechanism of the
nuclear disc. Orders of magnitude are consistent with the observations,
if about 25 clusters have been disrupted to form the nucleus,
which mass has grown as $t^{1/2}$. This means that one cluster is falling
every 400 Myr; since it will perturb the brightness distribution during
$\approx$ 0.2 Myr, the probability of the present configuration of the M31 
nucleus is about 4 10$^{-4}$. 
This however includes only globular clusters spiraling in on a 
quasi-circular orbit. We have shown that P1 being on a circular orbit
is inconsistent with the presence of a central dark mass of a few $10^7$~\Msun.
More rapid globular clusters, in hyperbolic
orbits, as in one of the scenarii proposed here, would be disrupted, but not
slowed down enough to contribute to the nucleus mass. 

Therefore, we should conclude that the hypothesis
of a falling cluster (e.g. nucleus of an accreted dwarf galaxy,
globular cluster) remains a viable explanation.
Lauer et al. (1993) suggested that P1 could survive 
close to the centre if it contains a secondary black hole
of ``substantial mass''. The probability of seeing it now in this 
configuration 
is even smaller than for a stellar cluster, given the frequencies
of black holes. Only if we assume
that dark matter in galactic halos is constituted by BHs are the latter
more numerous than globular clusters. In the model of
Lacey \& Ostriker (1985), one of these 10$^6$ M$_\odot$ BH arrives in
the centre every 10$^8$ yrs, and most of the time it merges with the BH
already present, or is ejected if there is already a binary BH. Xu \&
Ostriker (1994) argue that, since fresh BH arrive regularly on the centre,
the probability to find a binary BH with a separation of 1.6pc (as P1-P2) is
10\%. It is however difficult to explain why P1 is so bright, without any
cusp contrary to P2. More stringent is the constraint of a low dispersion
at P1: in the case of a secondary black hole at P1, the dispersion there should
be significantly larger than observed.

Our simulations finally served to precise the required BH mass, in 
the hypothesis of an axisymmetric nucleus: in order to
fit the disc thickness and stability requirements, we
need the presence of a mass concentration of 
at least $7\;10^7$~\Msun to explain the central stellar velocity and
velocity dispersion gradients. 
The detection of a strong central UV and X ray source in
the centre of M~31 is also in favor of a supermassive black hole.
The main underlying assumption in the kinematical determination
of the central dark mass is the {\em axisymmetry of the nucleus}.
In a forthcoming paper, we will examine the possibility
of a triaxial morphology (i.e. a bar) for the nucleus of M~31.

\begin{acknowledgements}
The authors would like to thank Roland Bacon
for fruitful discussions, Nicolas Cretton, Frank van den Bosch 
and Luis Aguilar for their comments and suggestions,
as well as Michael Rich for communication of data prior to publication.
Simulations have been carried out on the CRAY computers of the IDRIS-CNRS
center, in Orsay, France.
\end{acknowledgements}

\end{document}